\documentclass[letterpaper,twocolumn]{esapub} % American paper
\usepackage{cranm_timesv}
\usepackage{epsfig}
\usepackage{natbib}

\def\gtrsim{\mathrel{\hbox{\rlap{\hbox{\lower4pt\hbox{$\sim$}}}\hbox{$>$}}}}
\def\lesssim{\mathrel{\hbox{\rlap{\hbox{\lower4pt\hbox{$\sim$}}}\hbox{$<$}}}}
\newcommand{\para}{{\scriptscriptstyle \parallel}}

\title{\large
$\,\,\,\,\,\,\,\,\,\,$Why
is the Fast Solar Wind Fast and the Slow Solar Wind Slow?
\, \, \, \, \, \, \,
A Survey of Geometrical Models}

\author{Steven R. Cranmer}
\affil{Harvard-Smithsonian Center for Astrophysics,
Cambridge, MA 02138, USA \\
{\em Email:} scranmer@cfa.harvard.edu}

\begin{document}

% \keywords{coronal heating; MHD waves;
% solar corona; solar wind; turbulence; UV spectroscopy}

\maketitle

\begin{abstract}
Four decades have gone by since the discovery that the solar
wind at 1 AU seems to exist in two relatively distinct states:
slow and fast.  There is still no universal agreement concerning
the primary physical cause of this apparently bimodal distribution,
even in its simplest manifestation at solar minimum.  In this
presentation we review and extend a series of ideas that link the
different states of solar wind to the varying superradial geometry
of magnetic flux tubes in the extended corona.  Past researchers
have emphasized different aspects of this relationship, and we
attempt to disentangle some of the seemingly contradictory results.
We apply the hypothesis of Wang and Sheeley (as well as Kovalenko)
that Alfv\'{e}n wave fluxes at the base are the same for all flux
tubes to a recent model of non-WKB Alfv\'{e}n wave reflection and
turbulent heating, and we predict coronal heating rates as a
function of flux tube geometry.  We compare the feedback of
these heating rates on the locations of Parker-type critical
points, and we discuss the ranges of parameters that yield a
realistic bifurcation of wind solutions into fast and slow.
Finally, we discuss the need for next-generation coronagraph
spectroscopy of the extended corona---especially measurements
of the electron temperature above 1.5 solar radii---in order
to confirm and refine these ideas.
\end{abstract}

\vspace*{-7.05in}
\noindent
{\small
To be published in the proceedings of
{\em Solar Wind 11/SOHO--16: Connecting Sun and Heliosphere,}
June 13--17, 2005, Whistler, Canada,
ESA SP--592.}

\vspace*{6.33in}
\section{Introduction}

\vspace*{-0.033in}
The intertwined nature of solar wind acceleration and the
``coronal heating problem'' has been known since Parker (1958)
postulated a transonic flow solution made possible only by the
high gas pressure of the corona.
{\em Mariner 2} confirmed the existence of a continuous
supersonic solar wind in interplanetary space just a few years
after Parker's initially controversial work (for a first-hand
account of the discovery, see Neugebauer 1997).
{\em Mariner} also showed that the wind exists in two relatively
distinct states: slow (300--500 km/s) and fast (600--800 km/s).
The slow component was initially believed to be the
\linebreak[4]
``ambient'' background state (e.g., Hundhausen 1972), but it was
eventually realized that the fast component was in general more quiet
and steady (Feldman et al.\  1976; Axford 1977).
The polar passes of {\em Ulysses} in the 1990s confirmed this
revised paradigm (Gosling 1996; Marsden 2001).

In the 1970s and 1980s it became increasingly evident that
even the most sophisticated solar wind models could not
produce a {\em fast wind} without the deposition of
heat or momentum in some form into the corona
(e.g., Hartle and Sturrock 1968; Holzer and Leer 1980).
It also was realized that the geometry of the flow---i.e.,
whether the magnetic flux tubes were radially expanding
cones or superradially flaring trumpets---could have a
significant impact on the mass flux and wind speed.
This paper briefly surveys geometry-related explanations
for the observed distribution of solar wind speeds, and presents
a prediction for coronal heating rates in fast vs.\  slow
flux tubes as a consistency check on these ideas.

\vspace*{-0.033in}
\section{Coronal Source Regions}

\vspace*{-0.033in}
Even after several decades of ever-improving {\em in situ} and
remote-sensing observations, there is still no universal
agreement concerning the full range of coronal sources of the
solar wind.
It is clear that strong connections exist between large coronal
holes and the highest-speed wind streams (Wilcox 1968;
Krieger et al.\  1973; Noci 1973; Zirker 1977; see, however,
Habbal \& Woo 2001).
The more chaotic slow wind, though, may come from a multiplicity
of source regions.
Two regions that are frequently cited as sources of slow wind
are: (1) boundaries between coronal holes and large streamers
that undergo strong superradial expansion in the corona, and
(2) narrow plasma sheets that extend out from the tops of
streamer cusps (Wang et al.\  2000; Strachan et al.\  2002).
However, during active phases of the solar cycle, there is
evidence that slow wind also emanates from small coronal holes
(e.g., Nolte et al.\  1976; Neugebauer et al.\  1998)
and active regions (Hick et al.\  1995; Liewer et al.\  2004).
During the rising phase of solar activity, there seems to be an
abrupt ($\lesssim 6$ month) change in the magnetic connectivity
between field lines in the {\em in situ} ecliptic plane and the
Sun (see Figure 5 of Luhmann et al.\  2002).
At minimum, a large fraction of these field lines map into the
high-latitude northern and southern polar hole/streamer
boundaries, but at maximum nearly all the field lines map into
low-latitude active regions and small coronal holes.
The majority of the most recent transition time was
not observed by {\em SOHO} because of its 4-month mission
interruption in 1998.

The remainder of this paper is concerned mainly with
the dichotomy between high-speed wind that emerges from the
central regions of large coronal holes and low-speed wind
that emerges from the hole/streamer boundary regions.
More work is needed to apply the ideas presented below to other
potential slow-wind source regions (e.g., strong-field
flux tubes rooted in or near active regions;
see Wang 1994).

\vspace*{-0.07in}
\section{``Geometry is Destiny?''}

\begin{figure*}
\centering
\epsfig{figure=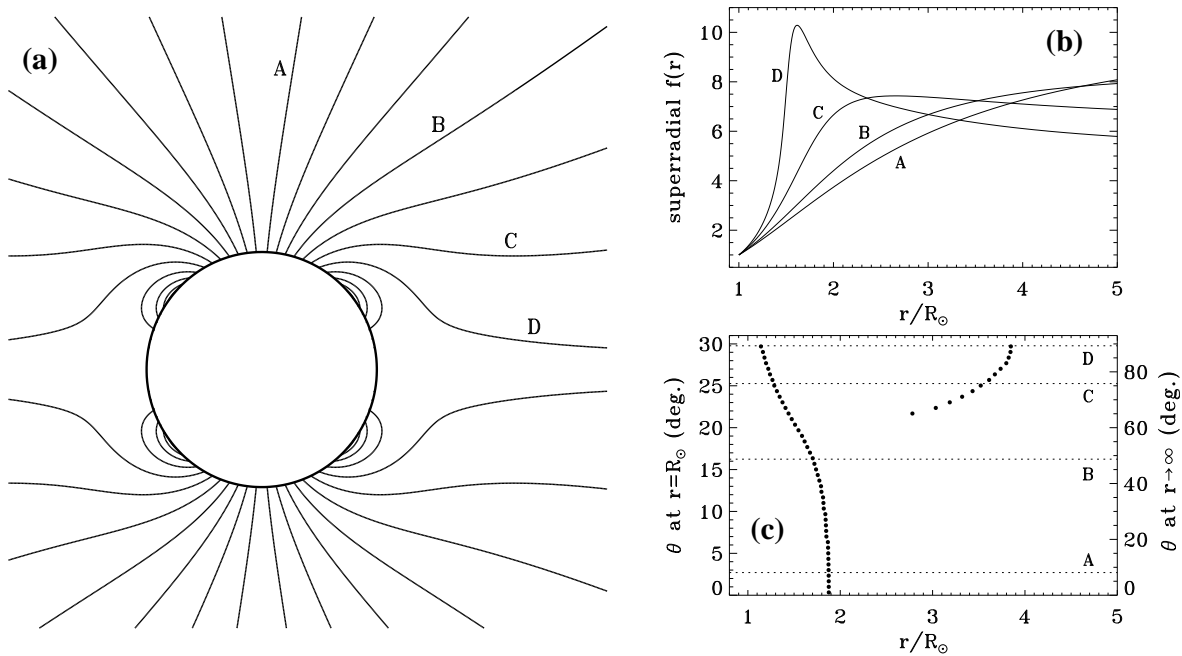,width=6.25in}
\caption{
{\bf (a)}
Idealized solar-minimum magnetic field configuration from the
model of Banaszkiewicz et al.\  (1998).
{Selected\-} field lines are labeled A $\! \rightarrow \!$ D in all 3
panels.
{\bf (b)}
Superradial flux-tube expansion factors (normalized to $f=1$ at
the solar surface) for 4 selected field lines.
{\bf (c)}
Possible radii of the sonic/critical point computed from local
minima in $F(r)$, shown for a fine grid of field lines in the
Banaszkiewicz et al.\  model (not all shown in [a--b]).
Colatitudes of field lines (measured from the pole) at $r=R_{\odot}$
and at infinity are plotted on the left and right vertical axes,
respectively.}
\end{figure*}

There is a strong empirical relationship between the solar wind
speed $u$ measured {\em in situ} and the inferred lateral expansion
of magnetic flux tubes near the Sun.
Levine et al.\  (1977) and Wang \& Sheeley (1990) found that
the asymptotic wind speed is inversely correlated with the amount
of transverse flux-tube expansion between the solar surface and
a reference point in the mid-corona (see also
Arge \& Pizzo 2000; Poduval \& Zhao 2004;
Fujiki, these proceedings).
As illustrated in Figures 1a and 1b, the field lines in the central
regions of coronal holes undergo a relatively slow and gradual
rate of superradial expansion, but the more distorted field lines
at the hole/streamer boundaries undergo more rapid expansion.
It should be noted, though, that the {\em eventual} flux tube
expansion (i.e., between the Sun and $r \!\rightarrow \!\infty$)
for polar coronal holes is likely to exceed that of the streamer
edges, despite the opposite trend seen when the expansion factor $f$
is measured between $R_{\odot}$ and a coronal source surface.

Several potential explanations for the observed anticorrelation
between wind speed and flux-tube expansion have been proposed
(see {\S}~4).
However, it is worthwhile to
\linebreak[4]
begin examining such a relationship
from the standpoint of the the equation of momentum
conservation along a solar wind flux tube:
\begin{equation}
  \left( u - \frac{a_{\para}^2}{u} \right)
  \frac{du}{dr} \, = \, \frac{dF}{dr}
\end{equation}
where, for a plasma dominated by protons and electrons, the
effective one-fluid most-probable speeds are defined as
$a^{2}_{\para / \perp} = k_{\rm B} (T_{p \, \para / \perp} + T_{e})/m_{p}$
and collisions and external sources of momentum are neglected.
The function $F(r)$ appearing on the right-hand side is defined as
\begin{equation}
  F(r) \equiv \frac{GM_{\odot}}{r} - a_{\para}^{2} +
  \int_{R_{\odot}}^{r} \! dr' a_{\perp}^{2} \left(
  \frac{2}{r'} + \frac{1}{f} \frac{df}{dr'} \right)
  \label{eq:Fofr}
\end{equation}
and $f(r)$ is the dimensionless flux-tube expansion factor (which
is proportional to $B^{-1} r^{-2}$ measured along a flux tube;
see also Kopp \& Holzer 1976).

Local extrema in $F(r)$ satisfy the Parker (1958) critical
point condition.
V\'{a}squez et al.\  (2003) found that only the
{\em global minimum} in $F(r)$ gives a sonic/critical point location
that allows a consistent and continuous solution for $u(r)$
over the full range of distances from the Sun to
\linebreak[4]
1 AU.
For monotonically increasing expansion factors like those over
the poles, $F(r)$ tends to exhibit a single minimum in the low
corona ($r \approx 2 \, R_{\odot}$).
For streamer-like expansion factors that peak near the cusp,
another minimum in $F(r)$ appears at a height well above the cusp;
this new point tends to be the global minimum.
The latter kind of flux tube---i.e., one that allows a more distant
critical point radius---seems to correspond directly to the
slow-speed wind measured {\em in situ} (see also
Wang 1994; Bravo \& Stewart 1997; Chen \& Hu 2002).

Figure 1c shows the radial locations of minima in $F(r)$ along
individually mapped flux tubes that range from the pole to the
edge of the streamer belt (see corresponding labels
A $\! \rightarrow \!$ D in the other panels).
Eq.~(\ref{eq:Fofr}) was solved using the magnetic field model
of Banaszkiewicz et al.\  (1998) and an isothermal corona
($T_{p \para} = T_{p \perp} = T_{e} =$
\linebreak[4]
1.75 MK) for simplicity.
The outer critical point appears only for field lines having
latitudes at $r \! \rightarrow \! \infty$ less than about
23$^{\circ}$ above and below the equator.
In more physically realistic models that include radial and
latitudinal temperature variations (e.g., V\'{a}squez et al.\  2003),
the outermost minimum in $F(r)$ is the global minimum, and thus
as one moves from the centers of coronal holes to their edges,
the critical point moves outwards {\em abruptly} from
$\lesssim 2 \, R_{\odot}$ to 3--6 $R_{\odot}$ at a latitude
still rather far removed from the streamer cusp.

\vspace*{-0.05in}
\section{Heating Above \& Below the Critical Point}

\vspace*{-0.07in}
Why does the height of the critical point matter?
Physically, the critical or singular point (equivalent to the
sonic point for a hydrodynamic pressure-driven wind) is the
location where the subsonic (i.e., nearly hydrostatic)
coronal atmosphere gives way to the kinetic-energy-dominated
supersonic flow.\footnote{%
The
\rule{0pt}{17pt}
idealized Parker critical point loses some of its mathematical
importance when solving the {\em time-dependent} momentum equation
(e.g., Suess 1982) or when including the effects of viscosity
(Axford \& Newman 1967).
However, the critical transonic `branch' remains the robust stable
time-steady solution in nearly all models with varying levels
of sophistication (e.g., Holzer \& Leer 1997; Velli 2001).}
Whether the critical point lies above or below the regions where
most of the energy deposition occurs is a key factor in
determining the nature of the wind:

\vspace*{-0.20in}
\begin{enumerate}
\item
If substantial heating occurs in the {\em subsonic} corona,
its primary impact is to ``puff up'' the scale height, drawing
more particles into the accelerating wind and thus increasing
the mass flux.
Roughly, the increase in energy flux due to the heating can be
balanced by the increase in mass flux, so that the eventual kinetic
energy per particle is relatively unaffected and the wind speed
may not change (relative to an unheated model).
In some scenarios the mass flux increase
can be stronger than the energy flux increase, and the asymptotic
wind speed decreases.
\item
If substantial heating occurs in the {\em supersonic}
corona, the subsonic temperature is unaffected and
the mass flux is unchanged.
The local increase in energy flux has nowhere else to go
but into the kinetic energy of the wind, and the flow speed
increases.
\end{enumerate}

\vspace*{-0.10in}
(Leer \& Holzer 1980; Pneuman 1980; Leer et al.\  1982).
The above dichotomy is often modeled by changing the height at
which the bulk of the energy is deposited, but it can also occur if
the heating remains the same and the height of the critical point
changes (as discussed in {\S}~3).

A natural link can be made between geometry-related changes in the
flow topology and the heating-related changes in the wind.
Wang \& Sheeley (1991) proposed that the observed anticorrelation
between $u$ and $f$ is a by-product of equal
amounts of Alfv\'{e}n wave flux emitted at the bases of
all flux tubes (see also earlier work by
Kovalenko 1978, 1981).
Near the Sun, the Alfv\'{e}n wave flux $F_A$ is proportional to
$\rho V_{A} \langle \delta V_{\perp} \rangle^{2}$.
The density dependence in the product of Alfv\'{e}n
speed $V_A$ and the squared Alfv\'{e}n wave amplitude
$\langle \delta V_{\perp} \rangle^{2}$ cancels almost
exactly with the linear factor of $\rho$ in the wave flux,
thus leaving $F_A$ proportional mainly to the radial
magnetic field strength $B$.
The ratio of $F_A$ at the critical point
to its value at the photosphere
thus scales as the ratio of $B$ at the critical point
to its value at the photosphere.
The latter ratio of field strengths is proportional to
$1/f$, where $f$ is the coronal expansion factor as
defined by Wang and Sheeley.
For equal wave fluxes at the photosphere for all regions,
coronal holes (with low $f$) will thus have a larger flux
of Alfv\'{e}n waves at and above the critical point compared
to streamers (that have high $f$).

To summarize, for streamers [coronal holes], more of the
Alfv\'{e}nic energy flux should be deposited
below [above] the critical point.
This effect is complementary to the change in height of
the critical point discussed above; i.e., for streamers
[holes] the critical point is high [low].

\vspace*{-0.01in}
\section{Turbulent Heating:$\,$ fast vs.\  slow}

\vspace*{-0.01in}
One aspect of the Wang/Sheeley/Kovalenko hypothesis that needs
further clarification is the link between an increased Alfv\'{e}n
wave flux and increased coronal heating.
Alfv\'{e}n waves can exert a dissipationless wave-pressure force
that can accelerate the wind (e.g., Isenberg \& Hollweg 1982),
but their ability to heat the plasma is less well understood. 
One idea that has received much recent attention is that
low-frequency Alfv\'{e}n waves can be damped in the
corona by undergoing a turbulent cascade from
large to small scales.
Here we present an empirically constrained model of Alfv\'{e}nic
turbulence and predict the contrast in extended heating that
occurs between a polar coronal hole flux tube and a
near-equatorial streamer edge flux tube.

Cranmer \& van Ballegooijen (2005) built a comprehensive model
of MHD turbulence in a polar coronal hole flux tube.
This model follows the radial evolution of the power spectrum of
non-WKB Alfv\'{e}n waves (i.e., waves propagating both outwards
and inwards along the flux tube) from the photosphere to 4~AU,
and allows the turbulent energy injection rate (and thus the
heating rate) to be derived as a function of height.
The Alfv\'{e}n waves have their origin in the transverse shaking of
strong-field ($\sim$1500 G) thin flux tubes in the photosphere.
The bottom boundary condition on the wave power spectrum
was derived from measurements of G-band bright point motions
in the photosphere (e.g., Nisenson et al.\  2003).
\linebreak[4]
Below the mid-chromosphere, where the bright-point flux tubes
are isolated and thin, the linear wave properties are computed
using a generalized form of the kink-mode wave equations derived by
Spruit (1981).
Above the mid-chromosphere, where the flux tubes have merged
into a more homogeneous network ``funnel'' (e.g.,
Tu et al.\  2005), the wind-modified non-WKB transport equations
of Heinemann \& Olbert (1980) are solved.

Figure 2 shows a summary of results from the published coronal-hole
model (corresponding to $\theta \! = \! 0$ in Figure 1) and the new
streamer-edge model (corresponding to field line D).
Below the transition region, the coronal-hole and streamer-edge
models were assumed to be identical except for the mass flux.
Figure 2a plots the adopted wind speed for both
flux tubes, as constrained by mass flux conservation, empirical
electron densities (e.g., Sittler \& Guhathakurta
1999) and the Banaszkiewicz et al.\  (1998) magnetic field model,
modified by including the thin tubes and funnels at low heights.
Also shown are the frequency-integrated Alfv\'{e}n wave amplitudes
$\langle \delta V_{\perp} \rangle$ as computed from the non-WKB
wave transport equations, with turbulent damping as described below.

\begin{figure}
\centering
\epsfig{figure=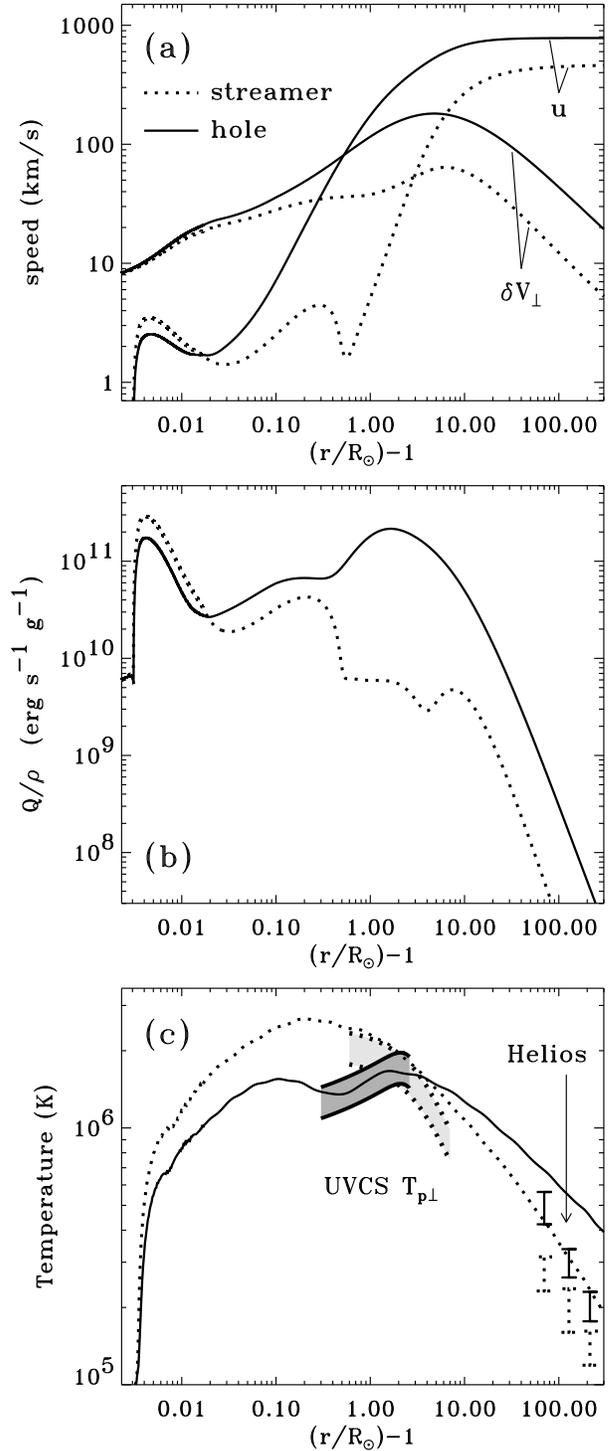,width=8.07cm}
\caption{
In all panels, solid lines represent the coronal-hole flux tube
(fast wind) and dotted lines represent the streamer-edge
flux tube (slow wind).
{\bf (a)} Outflow speed $u$ and 1D transverse Alfv\'{e}n wave
amplitude $\langle \delta V_{\perp} \rangle$ vs.\  height above
the photosphere.
{\bf (b)} Heating rates per unit mass.
{\bf (c)} Estimated one-fluid temperatures corresponding to the
heating rates in (b).  UVCS-derived proton temperatures and
in-situ Helios one-fluid temperatures are shown for comparison
(see text for references).}
\end{figure}

Figure 2b gives the derived heating rate---expressed as the energy
flux density $Q$ per unit mass density $\rho$---for the two
flux tube models.
The slowly-varying ratio $Q / \rho$ is plotted for convenience,
because $Q$ itself drops by more than 15 orders of magnitude between
the transition region ($r \approx 1.003 \, R_{\odot}$) and 1 AU.
In some ways, though, the plot is deceiving because it
seems as if the streamer flux tube is heated more than the
polar-hole model only below $r \approx 1.02 \,\, R_{\odot}$.
In fact, $Q_{\rm streamer} > Q_{\rm hole}$
\linebreak[4]
everywhere below
$r \approx 1.4 \,\, R_{\odot}$, but this is masked in the plot
because $\rho_{\rm streamer} > \rho_{\rm hole}$.
The heating rate is assumed to be equal to the
energy cascade rate that has been
derived for anisotropic MHD turbulence:
\begin{equation}
  Q \, = \, \rho \,
  \frac{\langle Z_{-}\rangle^{2} \langle Z_{+}\rangle +
        \langle Z_{+}\rangle^{2} \langle Z_{-}\rangle}
  {4 \ell_{\perp}}
\end{equation}
(Hossain et al.\  1995; Matthaeus et al.\  1999;
Dmitruk et al.\  2001, 2002),
where $\ell_{\perp}$ is a transverse outer-scale correlation length
and Elsasser (1950) variables are used to distinguish between
outwardly propagating waves ($Z_{-}$) and inwardly propagating
waves ($Z_{+}$), with
$Z_{\pm} \equiv \delta V \pm \delta B / \sqrt{4\pi\rho}$.
The correlation length is assumed to expand with the transverse
width of the flux tube (i.e.,
$\ell^{2}_{\perp} B = {\rm{const;}}$
see Hollweg 1986), and its normalization is specified by
Cranmer \& van Ballegooijen (2005).

Figure 2c shows the result of integrating a one-fluid internal
energy equation (e.g., eq.~3 of Leer \& Holzer 1980) to compute
the mean temperatures $T(r)$ that result from the heating rates
discussed above.
These results should be interpreted as preliminary because:
(1) they are not the result of a self-consistent calculation
of all fluid variables, and (2) a simple choice was made for
the electron heat flux (i.e., $q_{e \para} = q_{\rm SH} / 10$,
where $q_{\rm SH}$ is the classical Spitzer-H\"{a}rm value)
in order to split the difference between the known strong
conduction at low heights and the collisionless inhibition of
$q_{\para e}$ at large heights.
However, the overall trend in Figure 2c (i.e., the streamer
\linebreak[4]
being heated more at low heights, and less above the critical
point, than the coronal hole) is likely to remain valid when
the above approximations are corrected.

Several observations are also shown in Figure 2c, and the
general agreement in the hole/streamer contrast lends credence
to the overall validity of the approach outlined in this work.
UVCS/{\em{SOHO}} measured H~I Ly$\alpha$ resonance line profiles
in the extended corona which provide a probe of proton velocity
distributions.
For off-limb observations, the line of sight
samples directions that are mainly perpendicular to the $\sim$radial
field lines, and the $1/e$ line width $V_{1/e}$ arises from
two primary types of motion:
\begin{equation}
  V_{1/e}^{2} \, = \, \frac{2k_{\rm B} T_{p \perp}}{m_p} +
  \langle \delta V_{\perp} \rangle^{2}  \,\, .
\end{equation}
The two terms on the right represent random thermal
motions and unresolved transverse wave motions.
\linebreak[4]
Using measured values of $V_{1/e}$ and the modeled values of
$\langle \delta V_{\perp} \rangle$ for the waves,
the above equation was solved for the plotted $T_{p \perp}$.
The UVCS curve for a solar-minimum coronal hole (dark gray)
comes from an amalgam of data from Kohl et al.\  (1997),
Cranmer et al.\  (1999), Esser et al.\  (1999),
Zangrilli et al.\  (1999), and Antonucci et al.\  (2000).
The UVCS streamer data (light gray)
came from Kohl et al.\  (1997) and Strachan et al.\  (2004).
Also shown in Figure 2c are
{\em Helios} data points which were
computed by averaging together proton
(Marsch et al.\  1982) and electron (Pilipp et al.\  1990)
measurements, and isotropizing from bi-Maxwellian fits; i.e.,
taking $T \equiv (T_{\para} + 2 T_{\perp})/3$.

When examining predictions for the plasma temperatures in
fast vs.\  slow solar wind, it is worthwhile to compare with
different, but potentially complementary ideas.
Recently, Fisk (2003) and Schwadron \& McComas (2003) discussed
the origins of correlations between the eventual wind speed and
observed properties of emerging loops in the low corona (see also
the related footpoint diffusion model of Fisk \& Schwadron 2001).
Their prediction for there to be more {\em basal} coronal heating
(and a higher mass flux) in the slow wind seems to
be in accord with the results shown in Figure 2
(see also Matthaeus et al., these proceedings).
There still seems to be a disconnect, though, between theories
of coronal heating via flux emergence and theories that invoke
magnetic footpoint shaking (which in turn generates waves).
The relative contributions of these processes in various coronal
regions needs to be quantified further.

\section{Proton vs.\  electron heating}

The above analysis did not account explicitly for differences between
heating the various particle species in the plasma.
Even in a perfectly ``collisionally coupled'' plasma, there can
be {\em macroscopic} dynamical consequences depending on how the energy
is deposited into protons, electrons, and possibly heavy ions
as well.

Hansteen \& Leer (1995) demonstrated these effects for a
1D solar wind model:
when all of the heat goes into electrons, there is substantially
more downward conduction compared to a proton-heated model.
An electron-heated wind thus has a lower mid-corona temperature and
a lower wind speed than a proton-heated wind.
A 2D simulation of streamers by Endeve et al.\  (2004)
showed that the {\em stability} of closed-field regions is closely
related to this kinetic partitioning of heat.
When protons are heated strongly, the modeled
streamers become unstable to the ejection of massive plasmoids;
when the electrons are heated, the streamers are stable.
Any sufficiently predictive model of fast and slow solar wind 
must take these effects into account (see also
Cranmer \& van Ballegooijen 2003).

\section{Conclusions and Future Missions}

Our understanding of the dominant physics of solar wind
acceleration has progressed rapidly in the {\em SOHO} era, and
many of the insights embedded in the above analysis would not
have been possible without {\em SOHO.}
In particular, the strong preferential heating and acceleration
of heavy ions seen in coronal holes by UVCS has sharpened
theoretical efforts to understand kinetic energy deposition
in the collisionless extended corona (see reviews by
Axford et al.\  1999; Hollweg \& Isenberg 2002;
\linebreak[4]
Cranmer 2002a; Marsch 2004).

Despite these advances, the diagnostic capabilities of the
{\em SOHO} instruments were limited and the most fundamental
questions have not yet been answered.
If the kinetic properties of {\em additional ions} were to be measured
in the extended corona (i.e., a wider sampling of charge/mass
combinations) we could much better constrain the specific kinds of
waves that are present as well as the specific collisionless
damping modes (Cranmer 2001, 2002b).
Measuring the coronal {\em electron temperature} above
\linebreak[4]
$\sim$1.5 $R_{\odot}$
(never done directly before) would allow us to determine
the bulk-plasma heating rate in different solar wind structures,
thus putting the firmest ever constraints on models of why the
fast/slow wind is fast/slow.
Measuring {\em non-Maxwellian velocity distributions} of
electrons and positive ions would allow us to test specific models
of MHD turbulence, cyclotron resonance, and velocity filtration.
New capabilities such as these would be enabled by greater photon
sensitivity, an expanded wavelength range, and the use of measurements
that heretofore have only been utilized in a testing capacity
(e.g., Thomson-scattered H~I Ly$\alpha$ to obtain $T_e$).
Spectroscopy is key for the above measurements---especially in
combination with coronagraph occultation---in order to measure
detailed plasma properties out into the wind's acceleration region
(see also Kohl et al., these proceedings).

% \section*{Acknowledgments}

This work is supported by NASA
under grants NAG5-11913, NAG5-10996,
NNG\-04G\-E77G, and NNG\-04G\-E84G to the Smithsonian Astrophysical
Observatory, by Agenzia Spaz\-i\-ale Italiana, and by the
Swiss contribution to ESA's PRODEX program.

\end{document}